\documentclass[10pt,prd,aps,amssymb,amsmath,tightenlines,showkeys,twocolumn]{revtex4}

\usepackage{graphicx}

\usepackage[T1]{fontenc}
\usepackage[latin1]{inputenc}
\usepackage[english]{babel}
\usepackage{amsmath}
\usepackage{amsfonts}
\usepackage{amssymb}
\usepackage{dsfont}
\usepackage{color}

\usepackage[labelformat=parens]{subfig}
\usepackage[justification=raggedright,font=footnotesize]{caption}

\begin{document}

\title{Analytic eigenvalue structure of a coupled oscillator system beyond the ground state}

\author{Alexander Felski}\email{alexander-felski@t-online.de}
\author{S. P. Klevansky}\email{spk@physik.uni-heidelberg.de}

\affiliation{Institut f\"{u}r Theoretische Physik, Universit\"{a}t Heidelberg,  Germany}

\begin{abstract}

By analytically continuing the eigenvalue problem of a system of two coupled harmonic oscillators in the complex coupling constant $g$, we have found a continuation structure through which the conventional ground state 
of the decoupled system is connected to three other lower {\it unconventional} ground states that describe the different combinations of the two constituent oscillators, taking all possible spectral phases of these oscillators into account \cite{CHO}. In this work we calculate the connecting structures for the higher excitation states of the system and argue that - in contrast to  the four-fold Riemann surface identified for the ground state - the general structure is eight-fold instead. Furthermore we show that this structure in principle remains valid for equal oscillator frequencies as well and comment on the similarity of the connection structure to that of the single complex harmonic oscillator.

\end{abstract}


\maketitle
\section{Introduction}

For many years, studying functions of a complex variable has proved to be a useful - but very abstract - mathematical tool for many branches of physics. 
Recently, however, these sorts of techniques have taken on a new, important turn: first experiments using microwave cavities have shown that it is possible to circle a (square-root) exceptional (or branch) point, connecting the different Riemann sheets \cite{CD, BD}. Since then, further experimental studies have investigated the properties of parity-time ($\mathcal{PT}$) symmetric systems \cite{CMB}, detailing how one can move on the surfaces and around the exceptional points of these systems \cite{DOP, XU}.
Furthermore, a precise knowledge of the structure of the exceptional points has been shown to be important for designing tools for enhanced sensors \cite{CHEN}, making a knowledge of the analytic structure in the complex plane essential.

From a theoretical point of view, the structure of the Riemann surface is also interesting, not only because of the experimental possibilities that are opening up, but also because unexpected behavior can still be found:  In a recent study \cite{CHO} we examined the analytic eigenvalue structure of a 
system of two coupled harmonic oscillators described by the Hamiltonian 
{\small
\begin{equation}
\label{CHO_Hamiltonian}
H=p^2+\nu^2 x^2+q^2+\omega^2 y^2+g xy,
\end{equation}
}\noindent
where the $x$ and $y$ oscillators with natural frequencies $\nu$ and $\omega$ respectively are coupled linearly in both $x$ and $y$ through the coupling strength $g$. This system was shown to yield {\it four} possible ground state energies in the decoupling limit: $E_0(g=0)=\pm \nu \pm \omega$, a result which is obtained through continuing the complex coupling constant $g$ and assuming  that $\nu$ and $\omega$ are distinct \cite{CHO}. These states of the system describe the {\it conventional} ground state $E_0(g=0)=\nu + \omega$ as well as three other {\it unconventional} spectral phases of the two constituent oscillators, which are reached through analytic continuation.  This observation firstly suggests that states, which in principle have energy lower than the conventional ground state, can be reached by analytic continuation. Secondly, this brief analysis, together with an analysis of the first excited state in the decoupled theory appears to imply that this four-fold connection structure or Riemann surface structure is repeated for all excited states of the system.

In this paper, we question this implication, and analyse the excited states (including the first one) for general non-vanishing values of the complex coupling $g$ and show that the general Riemann surface structure is more intricate. 
In fact, we find that an eight-fold connection structure arises for a given total excitation of the system labelled by a single total quantum number $n$. In the decoupling limit, the spectrum consists of all possible combinations of the two oscillators that can give rise to $n$: if one oscillator has the quantum excitation $m$, the other necessarily has $n-m$. Thus, there is a symmetry under interchange of the oscillators. This makes it evident why the eight-fold Riemann structure collapses to a four-fold one, when, as for the ground state, each oscillator is forced to occupy the same excitation state. A regrouping of the energy eigenvalues into sets of four levels can be made in the decoupling limit; but knowing $E$ at this special value of the coupling, $g=0$, does not shed light on the full analytic structure of the Riemann surface.

In Sec.~IIA we review the calculation of the ground-state energy and visualize the four-fold Riemann surface. In Sec.~IIB, we move to the calculation of the first excited state, again providing a visualization of the Riemann surface and showing that this is eight-fold instead. We then generalize the method to calculate the energy function of $g$ for the $n$th existed state in Sec.~IIC, finding the eight-fold structure to be general. We compare the results from our asymptotic ansatz with those from a transformational one - while the latter seems easier to handle calculationally, it is not \emph{a priori} obvious that the transformational approach will account for the boundary conditions correctly and give the same result. This, however, turns out to be so. Finally, we re-examine the single complex harmonic oscillator in Sec.~IIIA and discuss the similarities to the coupled oscillator case in Sec.~IIIB. Some concluding remarks follow in Sec.~IV.

\section{The eigenvalue structure}
\subsection{Ground state energies}
\begin{figure*}
\centering
\subfloat[]{
\includegraphics[width=0.22\textwidth]
{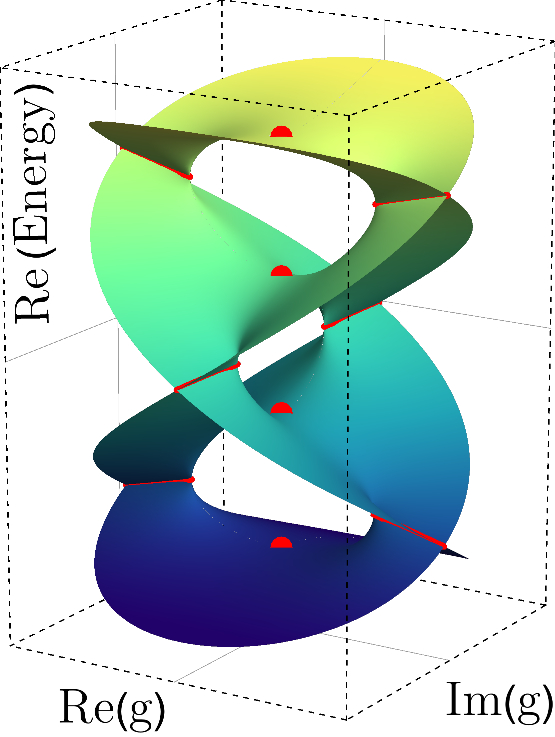}
\label{Riemann_0th_v1}
}
\quad
\subfloat[]{
\includegraphics[width=0.22\textwidth]
{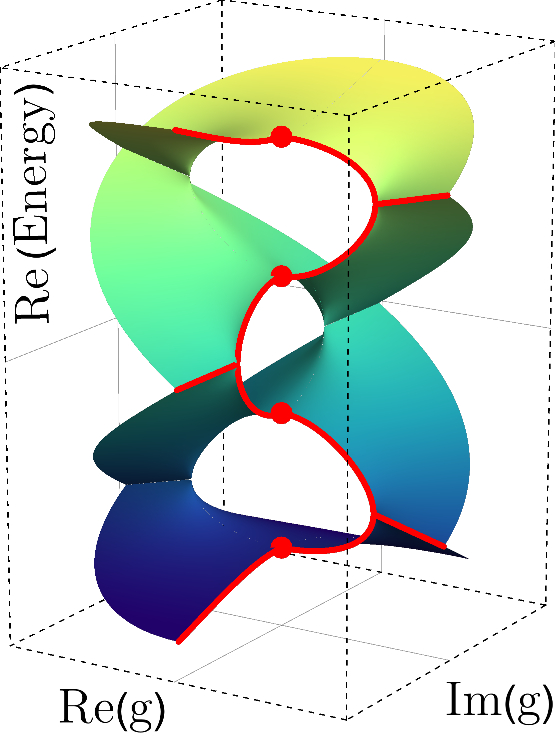}
\label{Riemann_0th_v2}
}
\quad
\subfloat[]{
\includegraphics[width=0.22\textwidth]
{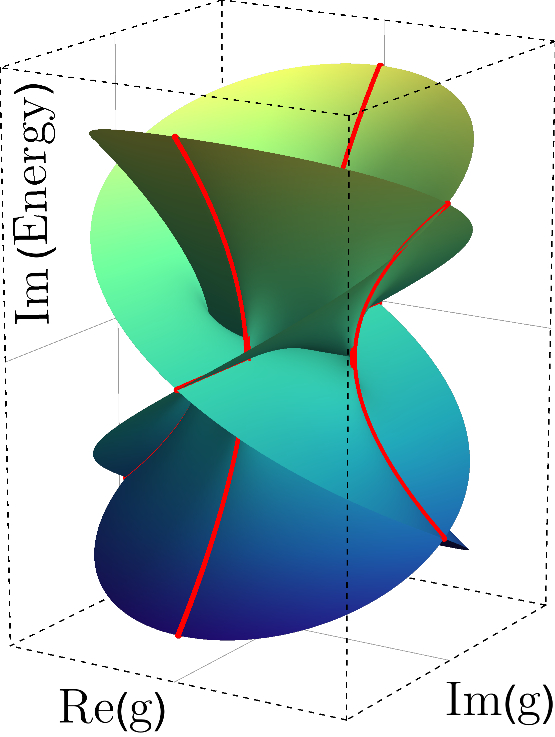}
\label{Riemann_0th_v3}
}
\quad
\subfloat[]{
\includegraphics[width=0.22\textwidth]
{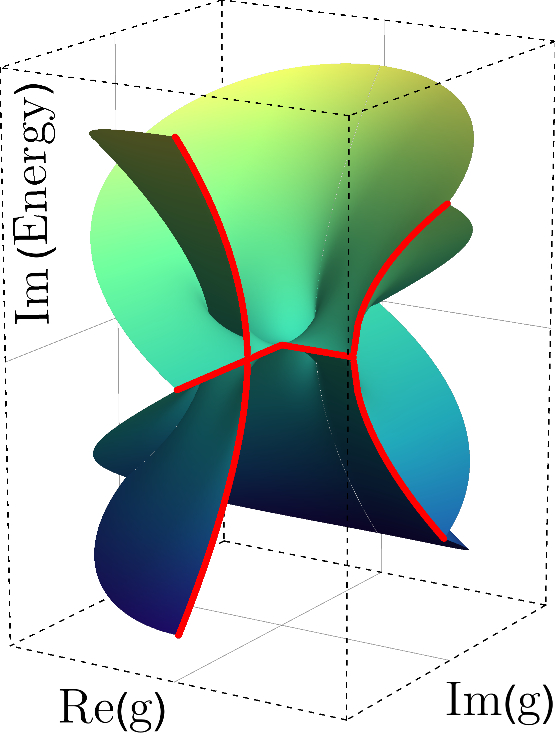}
\label{Riemann_0th_v4}
}
\caption[Riemann surface of the ground state of the coupled oscillator system 
over the complex coupling plane ($\nu>\omega$)]{Riemann surface 
structure of the ground-state energy function $E_0(g)$ over the complex 
coupling plane for $\nu>\omega$. Figures (a) and (b) show the real part of the 
Riemann surface. The red dots denote the four (purely real) decoupling limits 
$E_0(0)=\pm (\nu\pm \omega)$. In Fig.~(a) the branch cuts are denoted as solid red 
lines, while in Fig.~(b) the fourth quadrant of the coupling plane is cut out and the 
characteristics of the energy function at the cuts along the real and imaginary 
coupling axes are denoted as solid red lines. Figures (c) and (d) display this visualisation for 
the imaginary part of  $E_0(g)$.}
\end{figure*} 
In \cite{CHO} it was shown that with the ansatz 
{\small
\begin{equation}
\label{0state_wavefct}
\psi_0=a_{0,0} \cdot\mathrm{e}^{-\alpha x^2/2-\beta y^2/2 +\gamma xy}
\end{equation}
}\noindent
for the ground-state wave function, where the parameters $\alpha$, $\beta$ 
and $\gamma$ are to be determined and $a_{0,0}$ is an overall normalisation 
constant, Schr{\"o}dinger's eigenvalue equation $H \psi=E \psi$ for $H$ as given in (\ref{CHO_Hamiltonian}) leads to the system of equations
{\small
\begin{align}
\label{0state_energy}
E_0&=\alpha+\beta \\
\label{g_relation}
g\,&=-2(\alpha+\beta) \gamma \\
\label{nu_relation}
\nu^2&=\alpha^2+\gamma^2 \\
\label{omega_relation}
\omega^2&=\beta^2+\gamma^2
\, ,
\end{align} 
}\noindent
which describes the energy eigenvalues $E_0$ of the ground state and relates 
the parameters $\alpha$, $\beta$ and $\gamma$ to the natural oscillator 
frequencies and the complex coupling constant $g$. 
Eliminating these parameters in the energy equation (\ref{0state_energy}) leads 
to the quartic polynomial energy equation
{\small
\begin{equation}
E_0^4-2(\nu^2+\omega^2)E_0^2+(\nu^2-\omega^2)^2 +g^2 = 0
\end{equation}
}\noindent
that is solved by the ground-state energy function
{\small
\begin{equation}
\label{0state_energy_fct}
E_0(g)= \pm \sqrt{\nu^2+\omega^2 \pm \sqrt{4\nu^2\omega^2-g^2}}
\, .
\end{equation}
}\noindent
This nested square-root function describes a four-fold dependence of the energy on the complex coupling constant $g$. 
It has six (square-root type) branch points, four occurring at the coupling values $g=\pm 2\nu\omega$ and two at $g=\pm i \, (\nu^2-\omega^2)$ and the associated branch cuts allow the transition between the four sheets of the 
Riemann surface of $E_0(g)$ through a variation of the coupling constant in the complex plane. 
It was noted in particular that the possible ground states of the decoupled theory  $E_0(0)=\pm\nu\pm\omega$ correspond to the four possible combinations of the conventional and unconventional phases of the ground 
states of the two constituent oscillators and that the continuation in the coupling $g$ thus allows for a smooth transition between the states in this quartet structure. This result implies that one can in principle extract energy from the conventional ground state of the system. However, work must be done in moving $g$ in the complex plane to go from one sheet to another. One notes that for $g=0$ all possible values of $E_0(g)$ are purely real, a fact which can be traced back to the ${\mathcal{PT}}$ symmetry of the Hamiltonian for real values of $g$. In Figs.~\ref{Riemann_0th_v1}  and \ref{Riemann_0th_v3}  we have visualized the four-fold Riemann structure associated with this result. The red dots on these plots indicate the decoupling limit where $g=0$, while the branch cuts are shown as solid red lines. 
In Figs.~\ref{Riemann_0th_v2}  and \ref{Riemann_0th_v4} we have in addition cut out the fourth quadrant and indicated with a further solid red line a possible path that can be taken to move from one decoupling limit at $g=0$ to another in a continuous fashion.  

\subsection{First excited state}

In order to show that the quartet connection structure found for the ground state is repeated for higher excitation states of the system, one can investigate the structure of the decoupled theory for the first excited state of the system.  The goal is to show that the energy eigenvalues of the decoupled theory of the system can generally be collected in sets that contain four possible combinations of spectral phases of the excitation states of the constituent oscillators and that these four eigenvalues are analytic continuations of each other in the complex coupling plane. In \cite{CHO}, the {\it ad-hoc} ansatz for the wave function
{\small
\begin{equation}
\label{1state_wavefct_paper}
\psi=[a_{1,0}x+a_{0,1}y+a_{1,1}xy+a_{0,0}] \cdot  
\mathrm{e}^{-\alpha x^2/2-\beta y^2/2 +\gamma xy}
\end{equation}
}\noindent
was inserted into the eigenvalue equation and the resulting system of equations was analysed: it was found that the eigenvalues of the decoupled theory with $g=0$ can indeed be collected in such quartets. In the following we show, however, that already at this stage with the first excited state of the system, the analytic connection of the states is not restricted to these 
quartets and shows a more intricate structure.

In the discussion of the ground state the exponential wave function ansatz (\ref{0state_wavefct}) was chosen, as it is the most general function without roots that can satisfy the boundary conditions of an asymptotically vanishing wave function and balance the eigenvalue equation in the limit of large position variables $x$ and $y$.
For higher excitation states $\psi_n(x,y)$ the asymptotically dominating exponential behaviour remains, but instead of being multiplied with the single (normalising) overall constant $a_{0,0}$ as in (\ref{0state_wavefct}), we now include a 
polynomial prefactor $P_n(x,y)$ of $n$th degree to take into account the $n$ roots of the eigenfunction that we expect for the $n$th excited state of the system, in analogy to Courant's nodal line theorem that holds in the single 
position variable case:
{\small
\begin{equation}
\psi_n(x,y)= P_n(x,y)\cdot \mathrm{e}^{-\alpha x^2/2-\beta y^2/2 +\gamma xy}
\, ,
\end{equation}
where
}\noindent
{\small
\begin{equation}
P_n(x,y)= \sum_{i=0}^n \sum_{j=0}^i a_{i,i-j} x^i y^{i-j}
\, .
\end{equation}
}\noindent
For the first excited state this ansatz takes the form \footnote{Note that this ansatz differs from (\ref{1state_wavefct_paper}) in that the term $a_{1,1}$ does not appear. This coefficient was found in \cite{CHO} to be zero.}
{\small
\begin{equation}
\label{1state_wavefct}
\psi_1(x,y)= 
[a_{1,0} x + a_{0,1} y +a_{0,0}]\cdot 
\mathrm{e}^{-\alpha x^2/2-\beta y^2/2+
\gamma xy}
\end{equation}
}\noindent
Substituting it into Schr\"odinger's eigenvalue equation leads to the following system 
of equations
{\small
\begin{align}
\label{1state_E0}
0 &= a_{0,0}\, (E_1-\alpha\, -\beta) \\
\label{1state_g0}
0 &= a_{0,0}\,  (2(\alpha\, +\beta)\gamma+g)\\
\label{1state_nu0}
0 &= a_{0,0}\, (\alpha^2+\gamma^2-\nu^2)\\ 
\label{1state_omega0}
0 &= a_{0,0}\, (\beta^2+\gamma^2-\omega^2)\\[10pt]
\label{1state_Ecoupled1}
0 &= a_{1,0}\, (E_1-3\alpha\, -\beta) + a_{0,1}\, (2\gamma) \\
\label{1state_Ecoupled2}
0 &= a_{0,1}\, (E_1-\alpha\, -3\beta) + a_{1,0}\, (2\gamma) \\
\label{1state_gcoupled1}
0 &= a_{1,0}\, (\alpha^2+\gamma^2-\nu^2) - a_{0,1}\,  (2(\alpha\, +
\beta)\gamma+g)\\
\label{1state_gcoupled2}
0 &= a_{0,1}\, (\beta^2+\gamma^2-\omega^2) - a_{1,0}\,  (2(\alpha\, +
\beta)\gamma+g)\\
\label{1state_nu}
0 &= a_{0,1}\, (\alpha^2+\gamma^2-\nu^2)\\ 
\label{1state_omega}
0 &= a_{1,0}\, (\beta^2+\gamma^2-\omega^2)
\, .
\end{align}
}\noindent
We recognise the equations (\ref{1state_E0}) to (\ref{1state_omega0}) as being the ground-state system of equations (\ref{0state_energy}) to (\ref{omega_relation}) with the difference that the coefficient $a_{0,0}$ 
may vanish this time. However, if $a_{0,0}$ does not vanish the equations (\ref{g_relation}) to (\ref{omega_relation}) and  (\ref{1state_g0}) to (\ref{1state_omega0}) coincide, and the energy equations (\ref{0state_energy}) and (\ref{1state_E0}) 
imply that $E_1=E_0$. 
For the calculation of the general energy function $E_1(g)$ we can thus focus on the system of remaining equations (\ref{1state_Ecoupled1}) to (\ref{1state_omega}). This system again contains one part that describes the energy eigenvalues of the system and one that relates the parameters $\alpha$, $\beta$ and $\gamma$ to the natural oscillator frequencies $\nu$ and $\omega$ and the coupling constant 
$g$. The two coupled energy equations (\ref{1state_Ecoupled1}) and (\ref{1state_Ecoupled2}) can be combined to yield the quadratic energy equation
{\small
\begin{equation}
\label{1state_Edecoupled}
E_1^2-4(\alpha+\beta) E_1 +3(\alpha+\beta)^2+4(\alpha\beta-
\gamma^2)=0
\end{equation}
}\noindent
and the equations (\ref{1state_gcoupled1}) to (\ref{1state_omega}) are 
equivalent to the equations (\ref{g_relation}) to (\ref{omega_relation}) found in 
the ground-state system, because at least one of the coefficients $a_{0,1}$ and 
$a_{1,0}$ must not vanish in order for $P_1(x,y)$ to be a first degree polynomial. 
To eliminate the parameters $\alpha$, $\beta$ and $\gamma$ in terms of the parameters of the Hamiltonian $\nu$, $\omega$ and $g$ we use the definition of the ground-state energy (\ref{0state_energy}) and the relations
(\ref{g_relation}) to (\ref{omega_relation}) to rewrite equation (\ref{1state_Edecoupled}) as
{\small
\begin{equation}
\label{e1def}
E_1^2 - 4E_0 \cdot E_1+\Bigl(3E_0^2 \pm 2 \cdot 
\sqrt{4\nu^2\omega^2-g^2}\Bigr) = 0 
\, ,
\end{equation}
}\noindent
where the $\pm$ sign coincides with the choice of sign for the inner square root in $E_0(g)$. Now making use of the ground-state energy result (\ref{0state_energy_fct}), we can express the solutions to (\ref{e1def}) as
{\small
\begin{align}
\label{1state_energy_fct1}
\begin{split}
E_1^{+}(g) = 
\pm2&\sqrt{\nu^2+\omega^2+ \sqrt{4\nu^2\omega^2-
g^2}} \\
\pm&\sqrt{\nu^2+\omega^2-\sqrt{4\nu^2\omega^2-g^2}}  
\vspace{-.3cm}  
\end{split}
\end{align}
}\noindent
and
{\small
\begin{align}
\label{1state_energy_fct2}
\begin{split}
E_1^{-}(g) = 
\pm2&\sqrt{\nu^2+\omega^2- \sqrt{4\nu^2\omega^2-
g^2}} \\
\pm&\sqrt{\nu^2+\omega^2+\sqrt{4\nu^2\omega^2-g^2}}  
\, ,
\end{split}
\end{align}
}\noindent
taking care to choose the appropriate signs of the inner square root of $E_0(g)$. The possibilities $E_1^{+}(g)$ and $E_1^{-}(g)$, in which the inner square root of the first term in the solution is evaluated with a positive or negative sign respectively, are thus denoted separately.

\begin{figure*}
\centering
\subfloat[]{
\includegraphics[width=0.22\textwidth]
{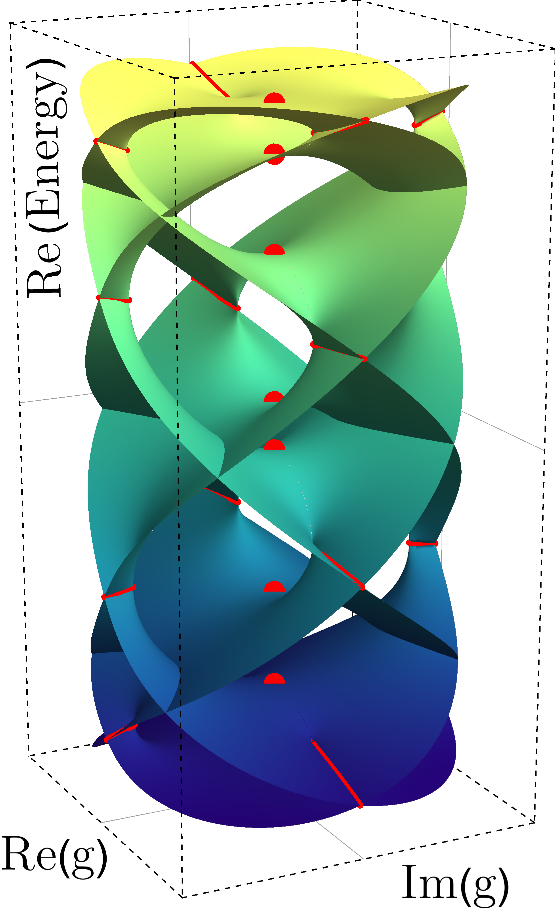}
\label{Riemann_1st_nu=4,omega=1_a}
}
\quad
\subfloat[]{
\includegraphics[width=0.22\textwidth]
{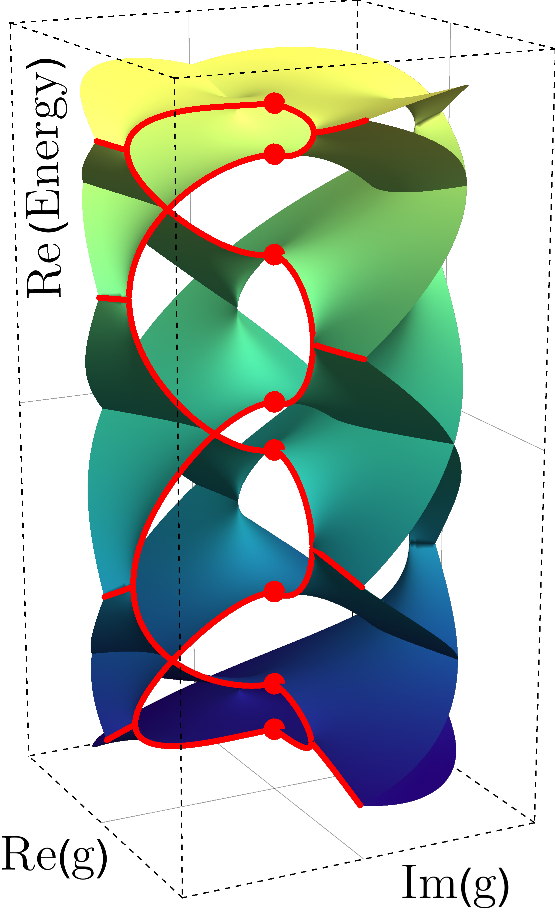}
\label{Riemann_1st_nu=4,omega=1_b}
}
\quad
\subfloat[]{
\includegraphics[width=0.22\textwidth]
{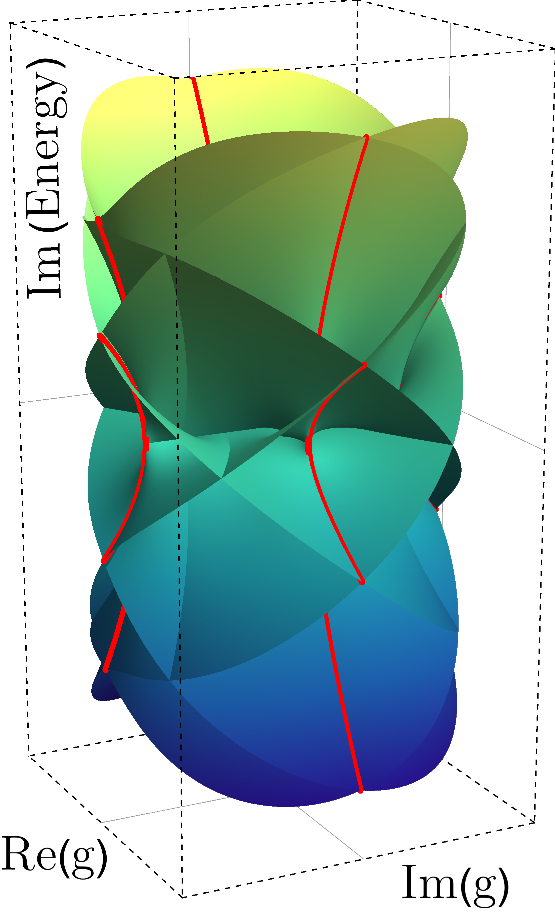}
\label{Riemann_1st_nu=4,omega=1_c}
}
\quad
\subfloat[]{
\includegraphics[width=0.22\textwidth]
{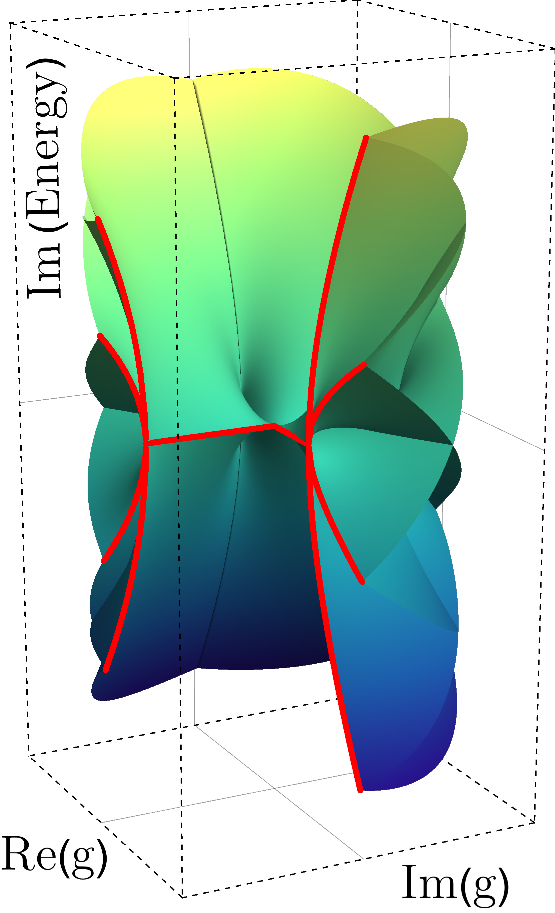}
\label{Riemann_1st_nu=4,omega=1_d}
}
\caption[Riemann surface of the first excited state of the coupled oscillator 
system over the complex coupling plane ($\nu>3\omega$)]
{Riemann surface structure of the first excited state energy 
function $E_1(g)$ over the complex coupling plane for $\nu>3\omega$. Figures 
(a) and (b) show the real part of the Riemann surface. The red dots denote the 
eight (purely real) decoupling limits $\pm(3\nu \pm \omega)$ and 
$\pm(\nu\pm3\omega)$. In Fig.~(a) the branch cuts are denoted as solid red lines, 
while in Fig.~(b) the fourth quadrant of the coupling plane is cut out and the 
characteristics of the energy function at the cuts along the real and imaginary 
coupling axes are denoted as solid red lines. Figures (c) and (d) display this visualisation 
for the imaginary part of  $E_1(g)$.}
\end{figure*} 

The eigenvalues of the decoupled theory of the first excited state can then be seen to be 
{\small
\begin{equation}
\label{1state_energy_fct1}
E_1^{+}(0) = \pm2(\nu+\omega) \pm(\nu-\omega)  
\vspace{-.3cm}  
\end{equation}
}\noindent
and
{\small
\begin{equation}
\label{1state_energy_fct2}
E_1^{-}(0) = \pm2(\nu-\omega) \pm(\nu+\omega)  
\, .
\end{equation}
}\noindent
These eight solutions can in fact be collected into two sets of four solutions: $\pm3\nu \pm\omega$ and $\pm\nu \pm3\omega$, which show a similar quartet structure to that found for the ground state of the system, recovering the result of \cite{CHO}. Each set contains the four possible combinations of conventional and unconventional phases of one constituent oscillator being in its first excited state and one being in its ground state. Although there is a resemblance to the analytic structure of the ground state, the general connection structure has changed: Instead of finding two separate quartets of analytically connected energies of the decoupled theory we find one single energy function $E_1(g)$ with sixteen (square-root type) branch points - eight at $g=\pm 2\nu\omega$ and eight at $g=\pm i \,(\nu^2-\omega^2)$ - with associated branch cuts that connect eight sheets of the Riemann surface of this energy function pairwise to one another. A visualisation of this surface is shown in Figs.~\ref{Riemann_1st_nu=4,omega=1_a} 
to \ref{Riemann_1st_nu=4,omega=1_d}.

The Riemann surface of the first excited state of the system thus describes an analytic connection of decoupled states in an octet rather than in a quartet. This octet contains the two quartets - each describing the four possible combinations of the conventional and unconventional spectral phases of the two constituent oscillators of the decoupled theory being in excited states whose respective quantum numbers add up to the total quantum number $n$ describing the excited state of the complete system - which are mirror images of each other under the exchange of the constituent oscillators in the decoupled theory. 

Furthermore, we can understand the four-fold structure of the ground state as a special case of this octet structure, in which the two quartets contained in the octet of decoupled states are identical because both constituent oscillators of the decoupled system are in the same excited state (namely the ground state).

\subsection{General excited states}

The analysis of the first excited state led us to the conclusion that the quartet structure of the energy levels that we found for the ground state of the system does not mean that the Riemann surfaces are four-fold for higher excited states. In particular, we found an octet structure for the first excited state that included the quartet structure as a special case ($g=0$). In this section we show that this octet structure of analytically connected energy levels of the decoupled 
system is in fact the fundamental structure of the system in the sense that it is repeated for higher excitations. 

Inserting the general wave function ansatz  
{\small
\begin{equation}
\label{nstate_wavefct}
\psi_n(x,y)= 
P_n(x,y)\cdot 
\mathrm{e}^{-\alpha x^2/2-\beta y^2/2+
\gamma xy}
\, ,
\end{equation}
}\noindent
into the eigenvalue equation and comparing coefficients in $x$ and $y$ leads 
to the set of equations
{\small
\begin{equation}
(M_n)_{j,i-j} =0\, ,
\label{eqM}
\end{equation}
}\noindent
where $i \in \{0,...,n+2\}$, $j\in \{0,...,i\}$ and $(M_n)_{j,i-j}$ is given as
{\small
\begin{align}
\label{M_j,i-j}
\begin{split}
(M_n)_{j,i-j} = \quad & \bigl [E_n -\alpha(2j+1) -\beta(2(i-j)+1) \bigr]
\,a_{j,i-j}  \\
+& \bigl[-(2(\alpha +\beta)\gamma+g) \bigr]\, a_{j-1,i-j-1}\\
+& \bigl[\alpha^2+\gamma^2-\nu^2 \bigr]\,a_{j-2,i-j}\\
+& \bigl[\beta^2+\gamma^2-\omega^2 \bigr]\,a_{j,i-j-2}\\
+& \bigl[(2\gamma)\cdot (j+1) \bigr]\,a_{j+1,i-j-1}\\
+& \bigl[(2\gamma) \cdot (i-j+1) \bigr]\,a_{j-1,i-j+1}\\
+& \bigl[(j+2)(j+1) \bigr]\,a_{j+2,i-j} \\
+& \bigl[(i-j+2)(i-j+1) \bigr]\,a_{j,i-j+2}
 \, .
 \end{split}
\end{align}
}\noindent
The polynomial factor $P_n(x,y)$ in the ansatz (\ref{nstate_wavefct}) is a polynomial of $n$th degree, such that at least one of the coefficients $a_{j,n-j}$ must not vanish. In particular, there exists one such coefficient with largest index 
$j$, such that $\forall k \in \{0,...,n\}$ with $k>j$ it follows that $a_{k,n-k}=0$. Using the equations $(M_n)_{j+2,n-j} =0$, $(M_n)_{j+1,n-j+1} =0$ and $(M_n)_{j,n-j+2} =0$, we conclude that the equations (\ref{g_relation}) to 
(\ref{omega_relation}), which relate the parameters $\alpha$, $\beta$ and $\gamma$ that determine the asymptotically dominating exponential behaviour of the eigenfunctions to the natural oscillator frequencies $\nu$ and $\omega$ 
and the coupling constant $g$, are generally satisfied for all excitations of the system.  Furthermore, this simplifies the expression (\ref{M_j,i-j}) such that the system of equations becomes a system of $(n+1)(n+2)/2$ coupled energy 
equations. The equations (\ref{eqM}) with $i=n$ form a subsystem which, after making the substitution $\lambda_k=\bigl[E_n -\alpha(2k+1) -\beta(2(n-k)+1) \bigr]$, takes the plain form
{\small
\begin{equation}
\label{eqs_antidiag.}
\begin{alignedat}{5}
&\bigl[M_n\bigr]_{0,n}:\quad &&\lambda_0\, a_{0,n}&&=\,\,&& 
1\cdot(-2\gamma)\,a_{1,n-1}\\
&\bigl[M_n\bigr]_{1,n-1}:\quad &&\lambda_1 \,a_{1,n-1}&&=\,\,&&
2\cdot(-2\gamma)\,a_{2,n-2}\\
& && && \,\,\,+&&n\cdot (-2\gamma)\,a_{0,n}\\
&\bigl[M_n\bigr]_{2,n-2}:\quad &&\lambda_2\, a_{2,n-2}&&=\,\,&& 
3\cdot(-2\gamma)\,a_{3,n-3}\\
& && && \,\,\,+&&(n-1)\cdot (-2\gamma)\,a_{1,n-1}\\
&\hspace{.25cm}...\\
&\bigl[M_n\bigr]_{n-1,1}:\quad &&\lambda_{n-1} \,a_{n-1,1}&&=\,\,&& 
n\cdot(-2\gamma)\,a_{n,0}\\
& && && \,\,\,+&&2\cdot (-2\gamma)\,a_{n-2,2}\\
&\bigl[M_n\bigr]_{n,0}:\quad &&\lambda_{n} \,a_{n,0} &&=\,\,&& 1\cdot 
(-2\gamma)\,a_{n-1,1}.
\end{alignedat}
\end{equation}
}\noindent
Solving this subsystem of equations leads to an expression for the energy eigenvalues: 
we combine the equations in (\ref{eqs_antidiag.}) by eliminating the 
coefficients $a_{l,n-l}$ to obtain a general polynomial energy equation for the $n$th excited 
state:
{\small
\begin{align*}
\prod_{i=0}^n \lambda_i=&- \hspace{-.3cm}\sum_{r=1}^{\left\lfloor (n+1)/
2 \right \rfloor}\biggl(\, \prod_{i=0}^n \lambda_i \biggr) \cdot \\
&\prod_{m=1}^{r}\, \biggl( \,\sum_{k_m=k_{m-1}+2}^{(n+1)-2(r+1-m)} 
\hspace{-.2cm} -\,\frac{(n-k_m)(k_m+1)(-2\gamma)^2}{\lambda_{k_m}\,
\lambda_{k_m+1}} \biggr) 
 \, .
\end{align*}
}\noindent
For the first five excited states this equation reads explicitly:
{\small
\begin{align*}
(n=0):\quad & E_0 -(\alpha + \beta) = 0\\[7pt]
(n=1):\quad &  E_1^2 -4 E_1\cdot(\alpha + \beta) +\\
&\Bigl(4(\alpha+
\beta)^2-((\alpha-\beta)^2+4\gamma^2)\Bigr) = 0\\[7pt]
(n=2):\quad & 
 E_2^3 -9 E_2^2\cdot (\alpha +\beta) +\\
 &E_2\Bigl(27(\alpha+
 \beta)^2-4((\alpha-\beta)^2+4\gamma^2)\Bigr) -\\
 &\Bigl(27(\alpha+
 \beta)^3-12(\alpha+\beta)((\alpha-\beta)^2+4\gamma^2)\Bigr)\\
 &= 0\displaybreak[1]
 \\[7pt]
(n=3):\quad & 
E_3^4 -16 E_3^3\cdot (\alpha +\beta) +\\
&E_3^2\cdot\Bigl(96(\alpha+
\beta)^2-10((\alpha-\beta)^2+4\gamma^2)\Bigr) -\\
&E_3\cdot
\Bigl(256(\alpha+\beta)^3-\\
&80(\alpha+\beta)((\alpha-
\beta)^2+4\gamma^2)\Bigr)+\\
&\Bigl(256(\alpha+\beta)^4-160(\alpha+\beta)^2((\alpha-
\beta)^2+4\gamma^2)+\\
&9((\alpha-\beta)^2+4\gamma^2)^2\Bigr)= 0
\\[7pt]
(n=4):\quad & 
E_4^5 -25 E_4^4\cdot (\alpha +\beta) +E_4^3\cdot\Bigl(250(\alpha+
\beta)^2-\\
&20((\alpha-\beta)^2+4\gamma^2)\Bigr) -\\
&E_4^2\cdot\Bigl(1250(\alpha+\beta)^3-\\
&300(\alpha+\beta)((\alpha-
\beta)^2+4\gamma^2)\Bigr)+\\
&E_4\cdot\Bigl(3125(\alpha+\beta)^4-\\
&1500(\alpha+\beta)^2((\alpha-
\beta)^2+4\gamma^2)+\\
&64((\alpha-\beta)^2+4\gamma^2)^2\Bigr)-\\
&\Bigl(3125(\alpha+\beta)^5-\\
&2500(\alpha+\beta)^3((\alpha-
\beta)^2+4\gamma^2)+\\
&320(\alpha+\beta)((\alpha-
\beta)^2+4\gamma^2)^2\Bigr)= 0
\end{align*}
}\noindent
and is solved by the following energy functions
{\small
\begin{align}
(n=0):\,\, &  
\label{m=0_ground_exc.}
E_0=1\cdot(\alpha \, + \beta)\, , 
\\[7pt]
(n=1):\,\, &  
\label{m=1_first_exc.}
E_1=2\cdot(\alpha \, + \beta)\pm 1\cdot \sqrt{ (\alpha \, -\beta)^2 
+4\gamma^2 } \, ,
\\[7pt]
(n=2):\,\, & 
\label{m=0_second_exc.}
E_2=3\cdot(\alpha \, + \beta)\, ,\\
\label{m=2_second_exc.}
&E_2=3\cdot(\alpha \, + \beta)\pm 2\cdot \sqrt{ (\alpha \, -\beta)^2 
+4\gamma^2 } \, ,
\\[7pt]
(n=3):\,\, & 
\label{m=1_third_exc.}
E_3=4\cdot(\alpha \, + \beta)\pm 1\cdot \sqrt{ (\alpha \, -\beta)^2 
+4\gamma^2 } \, ,\\
\label{m=3_third_exc.}
&E_3=4\cdot(\alpha \, + \beta)\pm 3\cdot \sqrt{ (\alpha \, -\beta)^2 
+4\gamma^2 } \, ,
\\[7pt]
(n=4):\,\, & 
\label{m=0_fourth_exc.}
E_4=5\cdot(\alpha \, + \beta) \, ,\\
\label{m=2_fourth_exc.}
&E_4=5\cdot(\alpha \, + \beta)\pm 2\cdot \sqrt{ (\alpha \, -\beta)^2 
+4\gamma^2 } \, ,\\
\label{m=4_fourth_exc.}
&E_4=5\cdot(\alpha \, + \beta)\pm 4\cdot \sqrt{ (\alpha \, -\beta)^2 
+4\gamma^2 } 
\, .
\end{align}
}\noindent
This is in particular in agreement with the results for the ground state (see Eq.~(\ref{0state_energy})) and the first excited state (see Eq.~(\ref{1state_Edecoupled})). We thus find that the energy function solutions take the form 
{\small
\begin{equation}
E_n=(n+1)\cdot (\alpha \, +\beta)\pm m\cdot \sqrt{ (\alpha \, +\beta)^2 
-4(\alpha\beta-\gamma^2) } 
\end{equation}
}\noindent
with $m$ running through all even natural numbers up to $n$ for even excitations $n$, 
i.e.  $m \in \{0\} \cup \{x\mid x \in \mathbb{N}\, \land\, (x \mid 2)\, \land\, (x\leq n) \}$, 
and $m$ running through all odd natural numbers up to $n$ for odd $n$, i.e.
$m \in \{x\mid x \in \mathbb{N} \,\land\, (x \nmid 2)\, \land\, (x\leq n) \}$. 
As in the discussion of the first excited state, this can be rewritten 
using the  result for the ground-state energy as
{\small
\begin{align}
\label{complete_coupled_energy}
\begin{split}
E_n^+(g)=
\pm (n+1)\cdot &\sqrt{\nu^2+\omega^2+
\sqrt{4\nu^2\omega^2-g^2}} \,\\
\pm m\cdot &\sqrt{ \nu^2+\omega^2-
\sqrt{4\nu^2\omega^2-g^2} } 
\vspace{-.1cm}
\end{split}
\end{align}
}\noindent
and
{\small
\begin{align}
\label{complete_coupled_energy2}
\begin{split}
E_n^-(g)=
\pm (n+1)\cdot &\sqrt{\nu^2+\omega^2-
\sqrt{4\nu^2\omega^2-g^2}} \,\\
\pm m\cdot &\sqrt{ \nu^2+\omega^2+
\sqrt{4\nu^2\omega^2-g^2} } 
\, .
\end{split}
\end{align}
}\noindent
We have again expressed the single energy function $E_n(g)$ in terms of two solutions 
$E_n^{+}(g)$ and $E_n^{-}(g)$, in order to emphasize the possible sign choices of the nested square-root functions.

The solutions in (\ref{complete_coupled_energy}) and (\ref{complete_coupled_energy2}) for arbitrary $n$ strongly resemble the energy function of the first excited state. And although they differ in the scaling factors of the two nested square-root functions, the eight-fold structure of the energy function solutions remains. The eigenvalues of the decoupled theory can now be collected into two quartets
$\pm (n+1+m)\, \nu \pm (n+1-m)\, \omega$ and 
$\pm (n+1-m)\, \nu \pm (n+1+m)\, \omega$ which are mirror images of each 
other under an exchange of the constituent oscillators of the decoupled theory; an analytic continuation of the coupling constant $g$ allows for a smooth transition between these eight states. Note as well that as $m$ runs through all non-negative even or odd numbers up to $n$ the coefficients of $\nu$ and $\omega$ in the two quartets run through 
the odd natural numbers up to $2n+1$ in such a way that the sum of the (absolute values of the) coefficients is $2(n+1)$, reflecting that for each possible value  of $m$ the sum of the quantum numbers of the constituent oscillators in the decoupled theory is the quantum number $n$ of the complete system, as  expected
If $m=0$ (see eg. (\ref{m=0_ground_exc.}), (\ref{m=0_second_exc.}) and (\ref{m=0_fourth_exc.})) we find four-fold energy functions which can be understood as special cases of the general octet structure in the way described in the last section: the two quartets, which are collected in an octet of decoupled states and which are mirror images of each other under an exchange of the constituent oscillators of the decoupled theory, are identical, because both constituent oscillators of the decoupled system are in the same excited state, i.e. the state with quantum number $n/2$.
The generally eight-sheeted Riemann surface of the energy function in (\ref{complete_coupled_energy}) and (\ref{complete_coupled_energy2}) simplifies to a four-fold structure (as that found in the discussion of the ground state of the system) accordingly, because the sheets coincide pairwise.

In contrast to the solutions that we found for the ground state and first excited 
state the solution for higher excited states cannot be written simply as a single analytic function.
The constant $m$ may take $\left \lfloor (n+1)/2 \right \rfloor$ different values 
at the $n$th excitation of the system. This means that although sets (quartets or octets) 
of energy eigenvalues of the decoupled theory at this excitation level are analytic continuations 
of each other, different sets remain separated. These different sets and their associated
Riemann surface can be characterised by the quantum number $n$ describing the 
excitation state of the complete system and the number $m$ describing the difference
of the excited states of the constituent oscillators in the decoupled theory.
We emphasise, in agreement with \cite{CHO}, that energy levels of different 
excited states of the complete system, i.e. belonging to sets with different values 
of $n$, are not analytically connected.
 
The energy functions in (\ref{complete_coupled_energy}) and 
(\ref{complete_coupled_energy2}) can furthermore be confirmed using canonical 
transformations of the Hamiltonian \cite{trafo} to decouple the system 
into the form 
{\small
\begin{equation*}
H' = p'^{\,2} + \Omega_+^2 \,x'^{\,2}+ q'^{\,2} + \Omega_-^2 \,y'^{\,2}\, .
\end{equation*}
}\noindent
with the effective oscillator frequencies
{\small
\begin{align*}
 \Omega_{\pm}= \sqrt{ \frac{(\nu^2+\omega^2)\pm \sqrt{\strut g^2 + (\nu^2 -\omega^2)^2 
 }}{2}  }\, .
\end{align*}
}\noindent
We note in particular that the wave functions resulting from this decoupled effective oscillator 
description satisfy the relations (\ref{g_relation}) to 
(\ref{omega_relation}) such that the boundary value structure of the 
transformed problem does indeed agree with the structure of the original 
system that we approached using a wave function ansatz motivated by 
an asymptotic analysis.
The transformational approach to solving the eigenvalue equation will, however, 
fail at the branch points of the energy function, because either the effective
frequencies vanish or the condition for the canonicity of the transformation is not
satisfied. 

\section{The structure of the analytic connection}

The connection of all four possible spectral phases of the two constituent oscillators
of the decoupled theory through an analytic continuation in the coupling constant $g$
is one of the principle results of \cite{CHO}. We showed that the four-fold connection structure of the energy that is found for the ground state does not repeat itself in all excited states of the complete system, as was
originally suspected in \cite{CHO}, but should instead be understood as a special case of an eight-fold structure that connects decoupled states, which are mirror images under an exchange of the constituent oscillators.
However, the result of an analytic connection between different spectral phases of the decoupled theory is retained. 

\subsection{Spectral phases of the harmonic oscillator}
\begin{figure}
\centering
\subfloat[]{
\includegraphics[width=0.173\textwidth]
{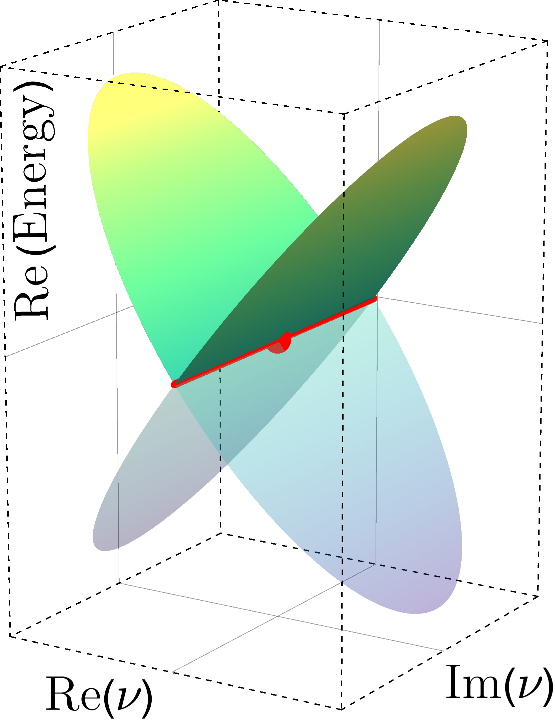}
\label{HO_Riemann_Re1}
}
\qquad
\subfloat[]{
\includegraphics[width=0.173\textwidth]
{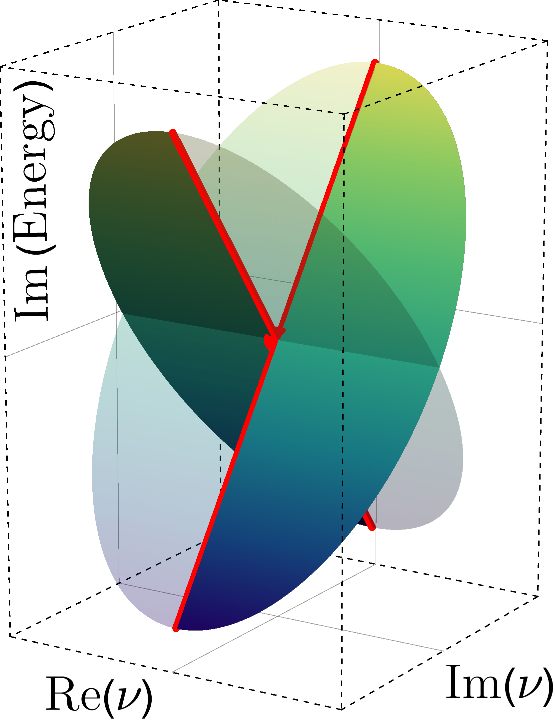}
\label{HO_Riemann_Im1}
}
\\
\subfloat[]{
\includegraphics[width=0.173\textwidth]
{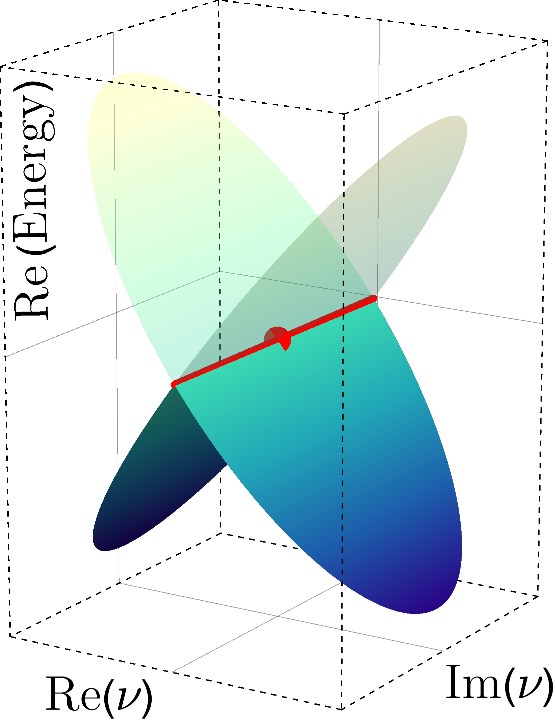}
\label{HO_Riemann_Re2}
}
\qquad
\subfloat[]{
\includegraphics[width=0.173\textwidth]
{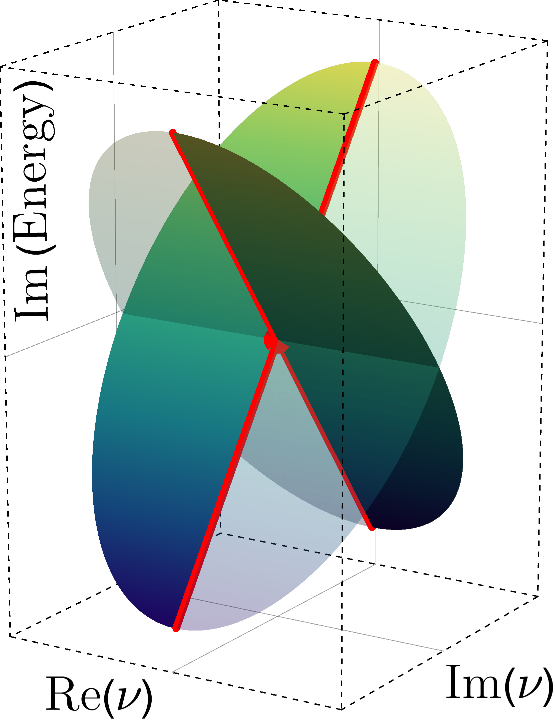}
\label{HO_Riemann_Im2}
}
\caption[Riemann surface of the harmonic oscillator ground state over the complex frequency plane.]{Riemann surface structure of the ground-state energy function $E_0(\nu)=\pm \sqrt{\nu^2}$ in the complex frequency plane. Figures (a) and (b) show the real and imaginary parts of the Riemann surface on the first sheet respectively. Figures (c) and (d) show the real and imaginary parts on the second sheet.  The dot in the center denotes the point $E_0(\nu=0)=0$, at which the two sheets coincide. This point can be characterised as two coalescing square-root branch points. The associated branch cuts are drawn as solid red lines along the imaginary frequency axis in the chosen representation. }
\end{figure} 
\begin{figure}
\centering
\subfloat[]{
\includegraphics[width=0.17\textwidth]
{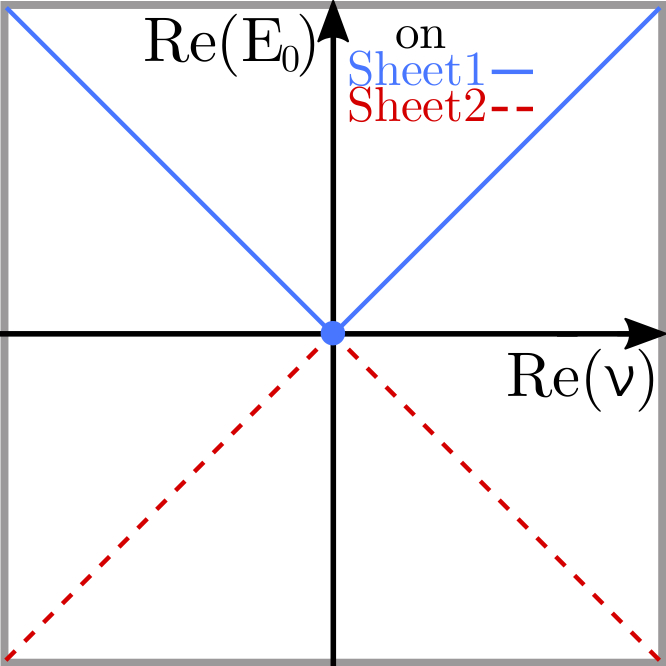}
\label{HO_linear_crossing_real_reE}
}
\qquad
\subfloat[]{
\includegraphics[width=0.17\textwidth]
{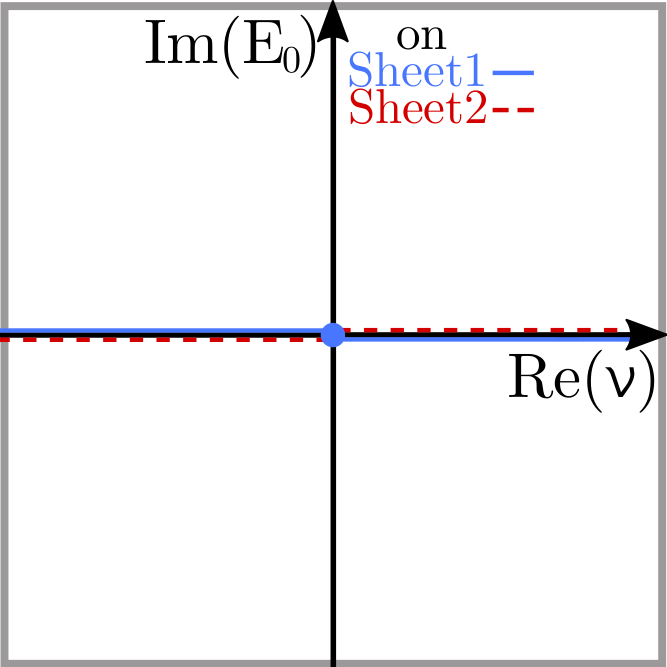}
\label{HO_linear_crossing_real_imE}
}
\\
\subfloat[]{
\includegraphics[width=0.17\textwidth]
{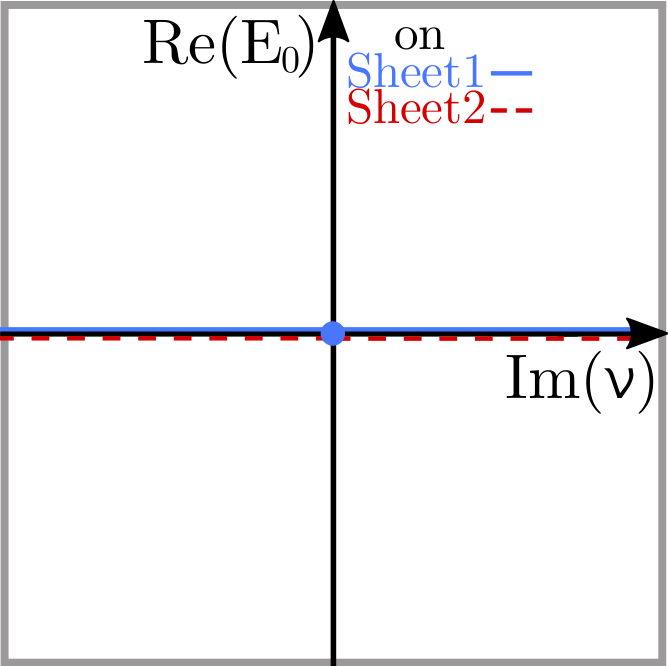}
\label{HO_linear_crossing_imag_reE}
}
\qquad
\subfloat[]{
\includegraphics[width=0.17\textwidth]
{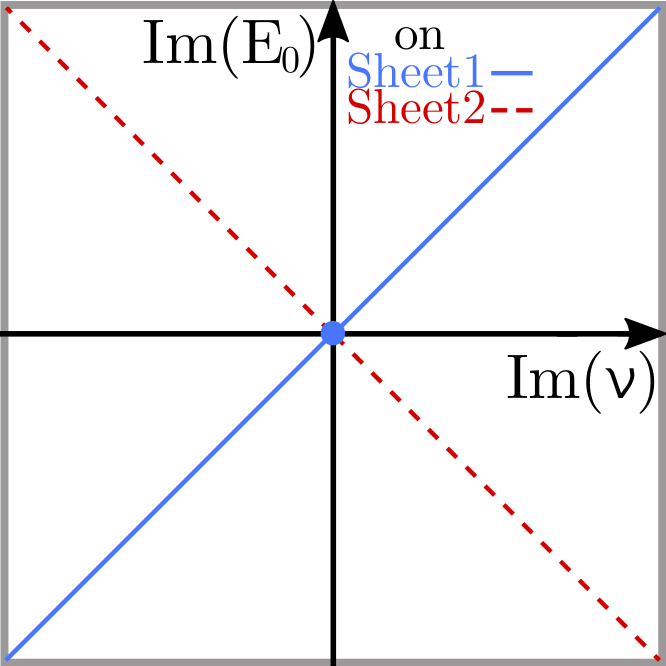}
\label{HO_linear_crossing_imag_imE}
}
\caption[Linear-crossing behaviour in the ground-state energy function of the harmonic oscillator]
{{
Behaviour of the real and imaginary parts of the harmonic oscillator ground state 
energy function $E_{0}(\nu)$ along the real and imaginary frequency axis. The 
behaviour on the first sheet is denoted by the solid blue line and on the second sheet by the dashed red one. 
The blue dot in the center of each graph denotes the coalescing square-root branch points at vanishing frequency.} }
\end{figure} 
The occurrence of different spectral phases and their connection through an analytic
continuation is in itself not surprising. A conventional and an unconventional 
spectrum can, for example, be found for the single harmonic oscillator, $H=p^2+\nu^2z^2$, where a connection 
between both is established by the analytic continuation of the natural oscillator 
frequency $\nu$ \cite{HO}. Accordingly, we can find basic structural properties of the connection structure of the 
coupled oscillator system within the system of a single harmonic oscillator.

As argued in \cite{HO} the ground-state energy of the (complex) harmonic oscillator takes
the form $E_0(\nu)=\pm \nu$, describing the conventional and the unconventional solutions.
Notice that the eigenvalues of the two different spectral phases of the solution only
agree for a vanishing frequency value. Although the energy function $E_0(\nu)$ can be 
calculated exactly in various ways, it is instructive to use a similar approach to the 
coupled oscillator system presented in Sec.~II. Inserting the ansatz 
{\small
\begin{equation}
\label{HO_wavefct_ansatz}
\psi_0(z,\alpha) = c \cdot \mathrm{e}^{-\alpha z^2/2}
\end{equation}
}\noindent
for the wave function into Schr{\"o}dinger's eigenvalue equation leads to the system of equations
{\small
\begin{align*}
E_0&=\alpha \\
\nu^2&=\alpha^2
\, ,
\end{align*} 
}\noindent 
which we can easily solve for the energy and obtain the expression $E_0(\nu)=\pm \nu$.
Nevertheless, the structure of the system of equations suggests that we should formulate
the energy function as $E_0(\nu)=\pm \sqrt{\nu^2}$. A Riemann surface representation 
for general complex frequency values is displayed in Figs.~\ref{HO_Riemann_Re1} to 
\ref{HO_Riemann_Im2}.
In the square-root notation $E_0(\nu)=\pm \sqrt{\nu}\cdot \sqrt{\nu}$ of the
energy function we notice that the solution has two coalescing branch points at $\nu=0$.
Furthermore, if we consider the behaviour of the energy eigenvalues as a function of 
purely real oscillator frequencies we find purely real eigenvalues that show a linear-crossing 
behaviour, which is similar to the behaviour of eigenvalues in the vicinity of a
diabolic point. In contrast to a typical diabolic point, at which the two 
eigenfunctions associated with the coalescing eigenvalues remain linearly independent,
the wave functions break down to the trivial solution in our case.
Moreover, if we consider the behaviour of the energy as a function of purely imaginary 
frequencies, we find purely imaginary eigenvalues that show the same linear-crossing
behaviour, see Figs.~\ref{HO_linear_crossing_real_reE} to \ref{HO_linear_crossing_imag_imE}.

\begin{figure}
\centering
\subfloat[]{
\includegraphics[width=0.18\textwidth]
{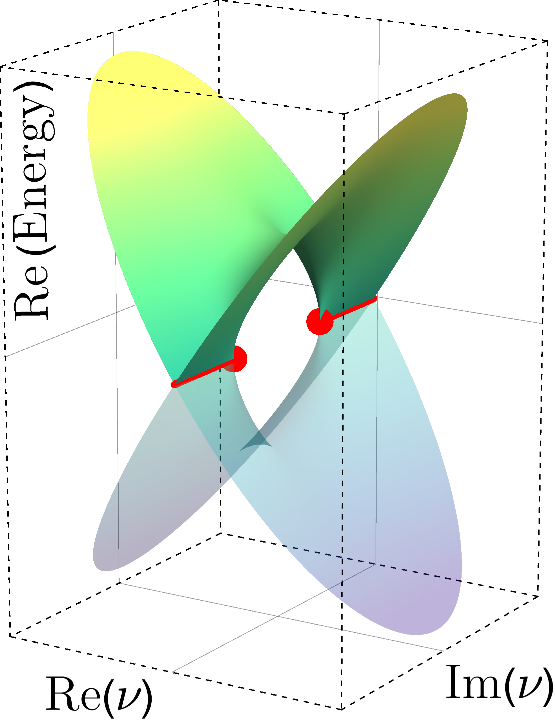}
\label{HOish2_Riemann_Re1}
}
\qquad
\subfloat[]{
\includegraphics[width=0.18\textwidth]
{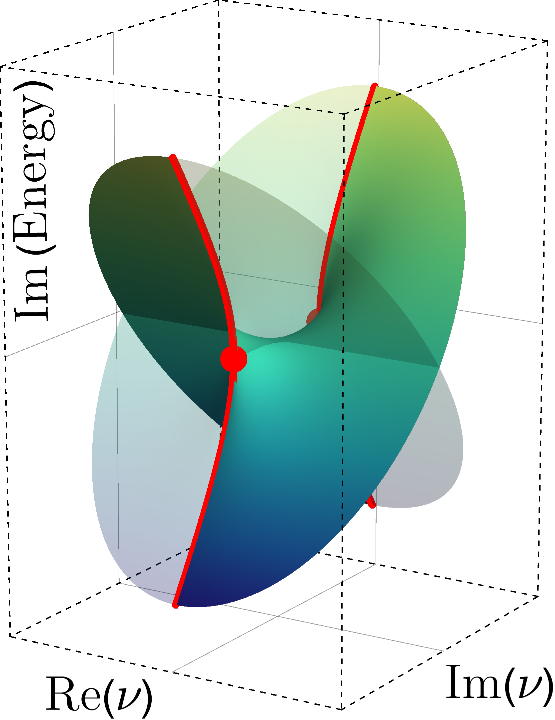}
\label{HOish2_Riemann_Im1}
}
\\
\subfloat[]{
\includegraphics[width=0.18\textwidth]
{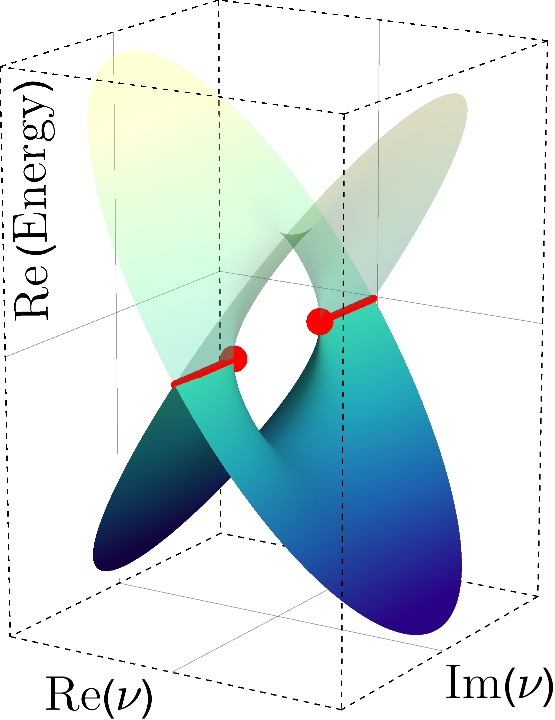}
\label{HOish2_Riemann_Re2}
}
\qquad
\subfloat[]{
\includegraphics[width=0.18\textwidth]
{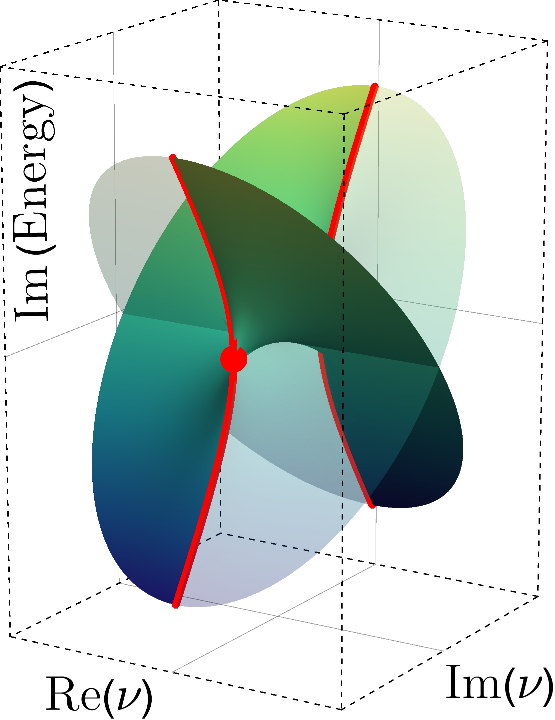}
\label{HOish2_Riemann_Im2}
}
\caption[Riemann surface of the modified harmonic oscillator ground state $E_0^+$ over the complex frequency plane.]{Riemann surface structure of the ground-state energy function $E_0^+(\nu)=\pm \sqrt{\nu-i\cdot\delta}\cdot \sqrt{\nu+i\cdot\delta}$ in the complex frequency plane. Figures (a) and (b) show the real and imaginary parts of the Riemann surface on the first sheet respectively, while Figs (c) and (d) show the real and imaginary parts on the second sheet.  The dots denote the positions of the square-root branch points at $\nu = \pm i \cdot \delta$. The associated branch cuts are drawn as red lines along the imaginary frequency axis. 
\label{HOish2_Riemann}
}
\end{figure}

The behavior of the coalescing square-root branch points appears similar to that which occurs at a 
diabolic point. This, as well as the similarity to the structure of the coupled oscillator system, becomes more apparent if we consider the following modification of the harmonic oscillator Hamiltonian:
{\small
\begin{equation*}
H^+ = p^2 +(\nu^2+\delta^2)\, z^2  \, ,
\end{equation*}
}\noindent
with some small real positive constant $\delta$.
For large frequencies this system resembles the common harmonic oscillator 
Hamiltonian. However, the wave function ansatz (\ref{HO_wavefct_ansatz}) leads to 
the ground-state energy function
{\small
\begin{equation*}
E_0^+(\nu)= \alpha = \pm \sqrt{\nu^2+\delta^2} = \pm \sqrt{\nu-i\cdot\delta}
\cdot \sqrt{\nu+i\cdot\delta}\, ,
\end{equation*}
}\noindent
which reveals the distinct behaviour at small frequencies: this function has two 
separate branch points on the imaginary frequency axis at $\nu=\pm i \cdot \delta$.
The associated Riemann surface is visualised in Figs.~\ref{HOish2_Riemann_Re1} to \ref{HOish2_Riemann_Im2}.
\begin{figure}
\centering
\subfloat[]{
\includegraphics[width=0.17\textwidth]
{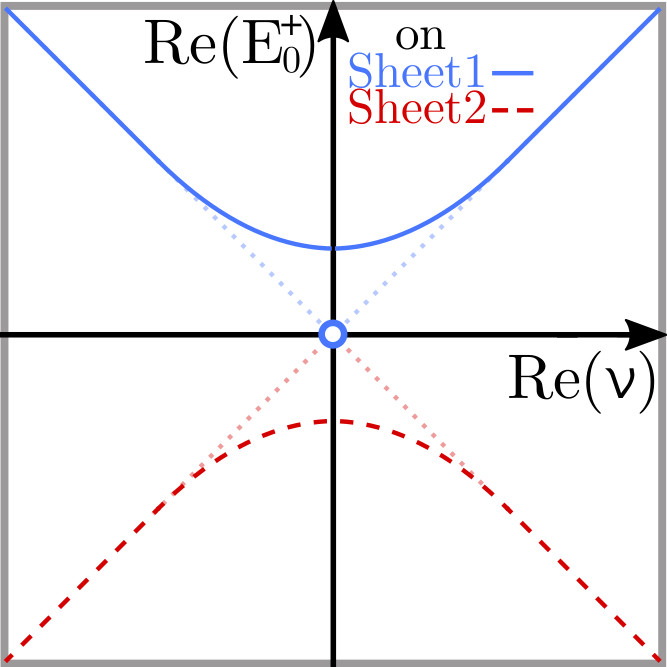}
\label{HOmod_linear_crossing_real_reE}
}
\qquad
\subfloat[]{
\includegraphics[width=0.17\textwidth]
{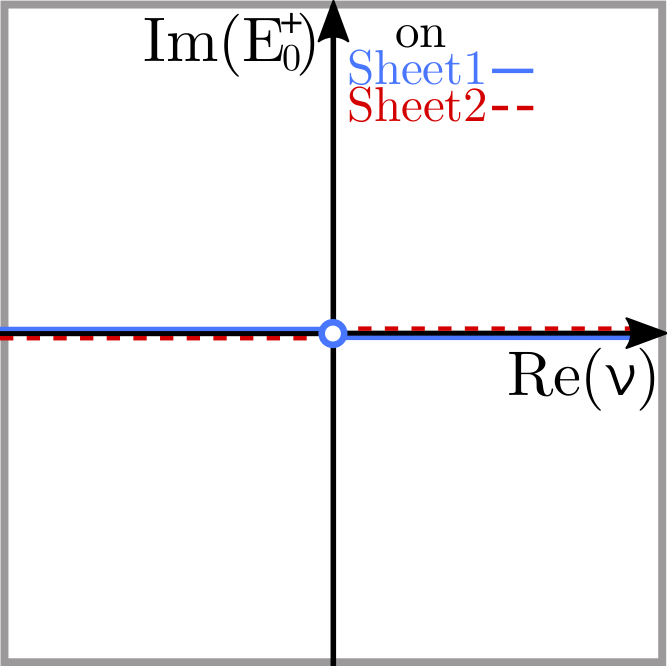}
\label{HOmod_linear_crossing_real_imE}
}
\\
\subfloat[]{
\includegraphics[width=0.17\textwidth]
{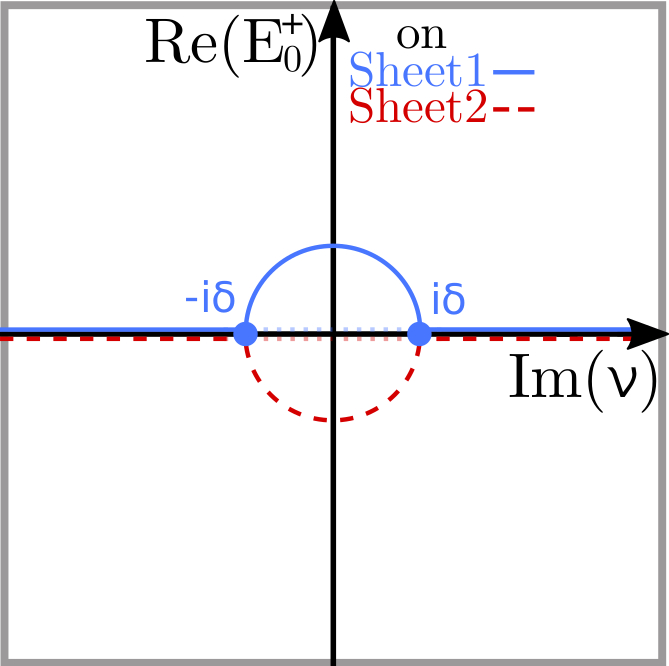}
\label{HOmod_linear_crossing_imag_reE}
}
\qquad
\subfloat[]{
\includegraphics[width=0.17\textwidth]
{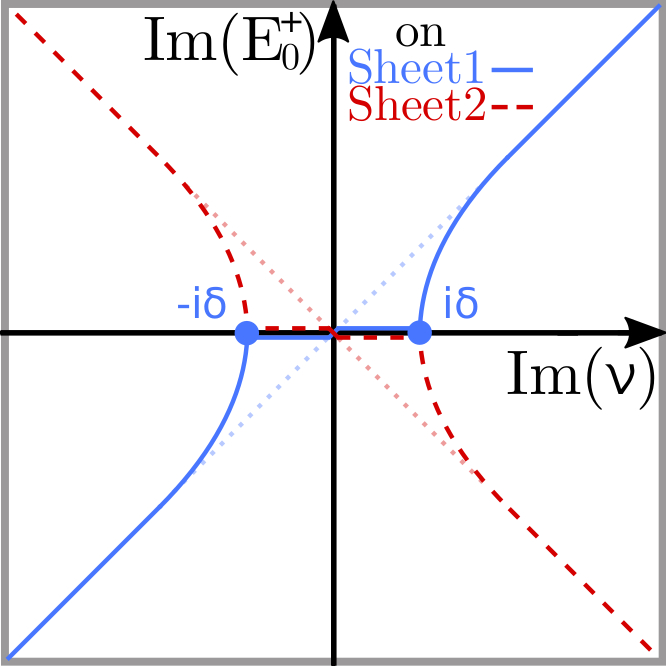}
\label{HOmod_linear_crossing_imag_imE}
}
\caption[Avoided-crossing behaviour and re-entrant symmetry in the ground-state 
energy function of the modified harmonic oscillator $H^+$]{
Behaviour of the real and imaginary parts of the modified harmonic oscillator 
ground-state energy function $E_{0}^+(\nu)$ along the real and imaginary frequency 
axis. The behaviour on the first sheet is denoted by the solid blue line and on the second sheet by the dashed red one. 
Solid blue dots denote the (imaginary) square-root branch points at $\nu=\pm i \cdot \delta$, 
while the open blue circle denotes the (vanishing) projection of these points onto the 
real frequency axis. }
\label{HOmod_linear_crossing}
\end{figure} 
For frequency values along the real frequency axis we still find purely real energy 
eigenvalues.
Notice, however, that the separation of the branch points along the imaginary frequency
axis introduces an upper or lower bound $\pm \delta$ on the energy eigenvalues at 
real frequency values; the linear-crossing behaviour of the harmonic oscillator 
becomes an avoided-crossing behaviour, see Figs.~\ref{HOmod_linear_crossing_real_reE} 
to \ref{HOmod_linear_crossing_real_imE}.
At the same time we notice that the eigenvalues associated with purely imaginary
frequencies no longer take on only imaginary values. For imaginary frequencies 
with $\lvert \nu \rvert <\delta$ we obtain real energies instead, see Figs.~\ref{HOmod_linear_crossing_imag_reE} to \ref{HOmod_linear_crossing_imag_imE}.
This behaviour has been referred to as \emph{re-entrant} $\mathcal{PT}$ symmetry
according to the symmetric nature of the Hamiltonian $H^+$ under parity-time reversal \cite{bubbles2}.
That author notes a resemblance to `bubbles of instability' in the equilibria of Hamiltonian systems,
as described in \cite{bubbles}. This resemblance is, however, superficial, as the bubble shown in Fig.~\ref{HOmod_linear_crossing} is associated with the existence of the two exceptional points. This differs from the result of \cite{bubbles}, which requires a nonconservative or dissipative perturbation of the Hamiltonian to arrive at such a structure \cite{ONK, ONK2}.
Note that the avoided-crossing behaviour along the real frequency axis as well as 
the the behaviour along the imaginary axis can generally be found in the study of 
diabolic degeneracies.
In \cite{diabolic} the eigenvalue surface near a diabolic point is unfolded due to a 
complex perturbation and the two possible structures found are the avoided crossing, 
which we encountered along the real frequency axis, and the so-called `double coffee filter'
behaviour. The latter is associated with the re-entrant symmetry and describes the behaviour
along the imaginary frequency axis in our system, where the roles of the real and imaginary 
energy contributions are exchanged compared to the behaviour along the real frequency axis.
Considering the (at least mathematical) connection of the two $\mathcal{PT}$-symmetric 
spectra of the modified oscillator potential that is represented in the Riemann surface 
in Fig. \ref{HOish2_Riemann} we may think of the missing real eigenvalues at small 
frequencies along the real frequency axis as being shifted onto the imaginary frequency axis, where they 
introduce a region of unbroken $\mathcal{PT}$ symmetry to the spectrum that 
does not occur in the harmonic oscillator system $H$.

The formation of a diabolic point in the harmonic oscillator limit $\delta\rightarrow 0$
can be confirmed utilising the fact that the behaviour in the vicinity of the separated
branch points is always equivalent to a two-dimensional problem, see \cite{chirality}.
For the system $H^+$ we find that the spectrum coincides globally with that of the 
matrix Hamiltonian 
{\small
\begin{equation*}
H^{\textrm{M}}=
\begin{pmatrix}
\delta & 0 \\
0 & -\delta
\end{pmatrix} 
+
\begin{pmatrix}
0 & \nu \\
\nu & 0
\end{pmatrix}
\, .
\end{equation*}
}\noindent
The eigenfunctions associated with the eigenvalues $E^{\textrm{M}}_{\pm}$ are
$\psi^{\mathrm{M}}_+=\big(\delta/\nu-\sqrt{1-(\delta/\nu)^2}\, ,\, 1\big)^T$ and 
$\psi^{\mathrm{M}}_-=\big(\delta/\nu+\sqrt{1-(\delta/\nu)^2}\, ,\, 1\big)^T$. 
If the frequency approaches the branch points $\nu \rightarrow \pm i \, \delta$, 
the eigenvectors of both sheets approach the same structure 
$\psi^{\mathrm{M}}_{+}, \, \psi^{\mathrm{M}}_{-} \rightarrow (\mp i \, , \, 1)^T$.
This reflects the exceptional point nature of the square-root branch points.
Furthermore, we can assign a definite left or right chirality to the branch points at 
$\nu = \pm i \, \delta$ depending on the structure of the eigenfunctions at those points, 
 \cite{chirality}, and notice that the two branch points have in any case opposite chiralities.
On the other hand, if we consider the general eigenvector structure in the limit of 
coalescing branch points, i.e. when $\delta \rightarrow 0$, we find that the eigenvectors 
take the form 
$\psi^{\mathrm{M}}_{+} \rightarrow (-1 \, , \, 1)^T$ and 
$\psi^{\mathrm{M}}_{-} \rightarrow (+1 \, , \, 1)^T$. 
They are independent of the frequency and in particular remain linearly independent 
at the point $\nu=0$ at which the two branch points coalesce.
Therefore the energy function of the harmonic oscillator system does indeed have a diabolic 
point at vanishing frequency $\nu$.
This behaviour of two coalescing branch points with opposite chiralities forming a diabolic 
point has been described and measured by K. Ding et al. in a system with a four-state non-Hermitian Hamiltonian with coupling \cite{coalescence}.
The breakdown of the eigenfunctions $\psi_0(z)$  that we found in the study of the 
harmonic oscillator system is an independent phenomenon that occurs, because for 
vanishing frequencies the Hamiltonian $H$ takes the form of the free-particle Hamiltonian. 
The eigenfunctions should therefore take the form of plane waves. However, the 
boundary condition of vanishing wave functions imposed in the large position limit require these 
oscillating plane waves to vanish identically, causing the eigenfunction solution to break 
down. 

\subsection{The coupled system at equal frequencies}

In this subsection we return to the analysis of the coupled harmonic oscillator system described by the Hamiltonian in (\ref{CHO_Hamiltonian}), and view its structure in the light of the knowledge gained in Sec.~IIIA.
In the following we show that we can, in fact, find similar properties 
in the connection structure of the coupled system: a connection to 
$\mathcal{PT}$-symmetric cases of the system and the formation of diabolic points from coalescing 
branch points.

The decoupled theory of (\ref{CHO_Hamiltonian}) describes a 
$\mathcal{PT}$-symmetric system with purely real eigenvalues, indicating an
unbroken symmetry phase.
The introduction of a purely real non-vanishing coupling constant $g$ preserves this
symmetry and the energy eigenvalues remain real in a weakly coupled system.
The unbroken symmetry phase for small coupling values does, however, break down 
when the coupling reaches the two (square-root) singularities of the general energy 
functions at $g=\pm 2\nu\omega$. For larger coupling values the eigenvalues become 
complex reflecting that the system is in a region of broken $\mathcal{PT}$ symmetry.
The  Hamiltonian (\ref{CHO_Hamiltonian}) is, however, only $\mathcal{PT}$-symmetric 
in the special case of purely real coupling values and the analytic continuation in the 
coupling is foremost a mathematical tool to study the connection of the different
spectral phases of the system. 
Nevertheless, real eigenvalues can not just be found in the unbroken symmetry region 
of real coupling values; we find them for purely imaginary couplings as well.
At purely imaginary coupling values up to the general (square-root) singularities of the 
energy functions at $g = \pm i \, (\nu^2-\omega^2)$ the eigenvalues are real and 
beyond these points they become complex. An exception is formed by energy levels belonging
to quartet connection structures, where the energy remains real on the first and fourth
sheet of the associated Riemann surface (corresponding to both constituent oscillators 
being in their conventional or both being in their unconventional phase in the decoupled
limit) even beyond these points. 
This behaviour was first described in \cite{partial} and related to the \emph{partial}
$\mathcal{PT}$ symmetry of the Hamiltonian, in which parity reflection only acts on 
one of the two position variables $x$ and $y$.
We thus find that the characteristic connection structure of energy levels through 
an analytic continuation in the coupling constant $g$ does not only enable the smooth 
transition between energy levels of different spectral phases,
but that the square-root singularities of the energy functions, which essentially
determine this structure, are related to the behaviour in the (partial) 
$\mathcal{PT}$-symmetric cases of the Hamiltonian as well.
In these cases the the energy functions generally show an avoided-crossing behaviour
along one coupling constant axis and an re-entrant behaviour along the other axis, similar
to that found for the single harmonic oscillator.

Furthermore, the two imaginary branch points in the energy function of the coupled system
at $g=\pm i \cdot (\nu^2-\omega^2)$ coalesce in the limit of equal oscillator frequencies. The avoided-crossing behaviour becomes a linear-crossing behaviour of the energy levels (and the re-entrant bubble vanishes).
At distinct natural oscillator frequencies $\nu$ and $\omega$ we can utilize the fact that the behaviour in the vicinity of the separated branch points  is always equivalent to a two dimensional problem: we can assign a chirality to the branch points and find that the two imaginary branch points always have the opposite chirality. Thus two imaginary square-root branch points coalesce for the decoupled theory at $g=0$ and form a diabolic point similar to that found for the single harmonic oscillator. This also implies that the transformational approach fails for the 
decoupled theory. But the approach using a  wave function ansatz based on asymptotic analysis is applicable and leads to the expected result, in which we simply take the equal frequency limit of the energy function solutions found earlier. There is, however, one special case: for vanishing coupling and equal oscillator frequencies $\nu=\omega$ the relations 
(\ref{g_relation}) to (\ref{omega_relation}) imply that $\alpha=\beta=\pm \nu$ or $\alpha=-\beta=\pm \nu$.
\begin{figure*}
\centering
\subfloat[]{
\includegraphics[width=0.22\textwidth]
{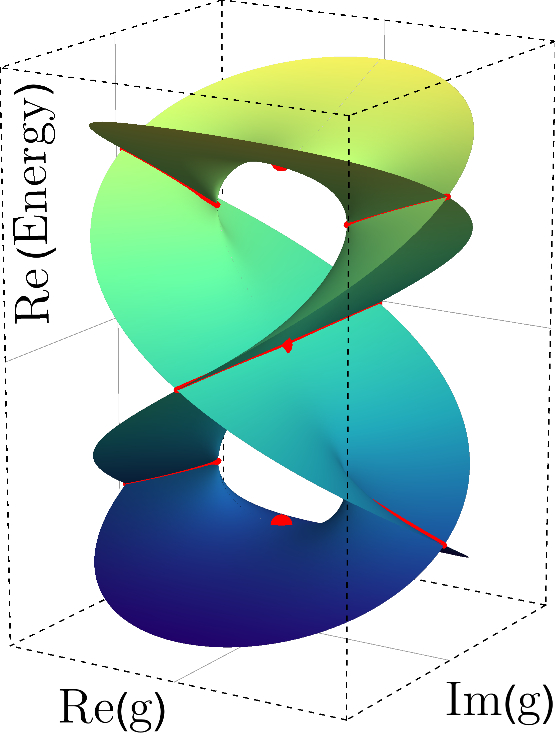}
\label{Riemann_0th_nu=omega_v1}
}
\quad
\subfloat[]{
\includegraphics[width=0.22\textwidth]
{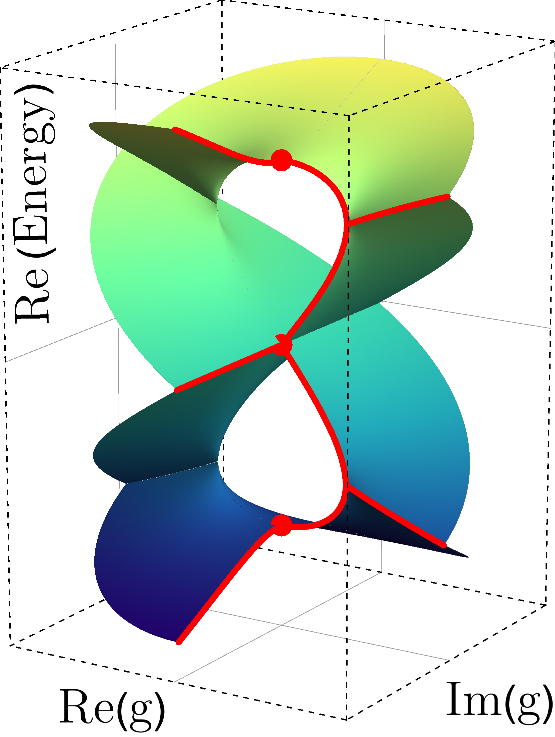}
\label{Riemann_0th_nu=omega_v2}
}
\quad
\subfloat[]{
\includegraphics[width=0.22\textwidth]
{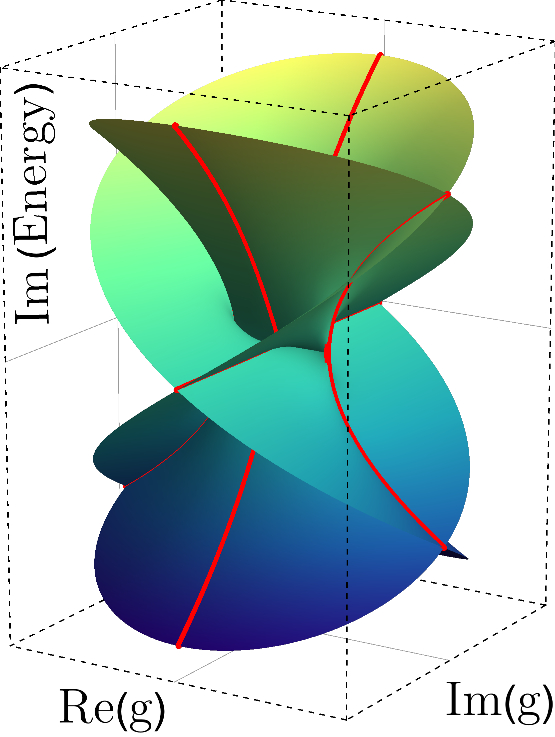}
\label{Riemann_0th_nu=omega_v3}
}
\quad
\subfloat[]{
\includegraphics[width=0.22\textwidth]
{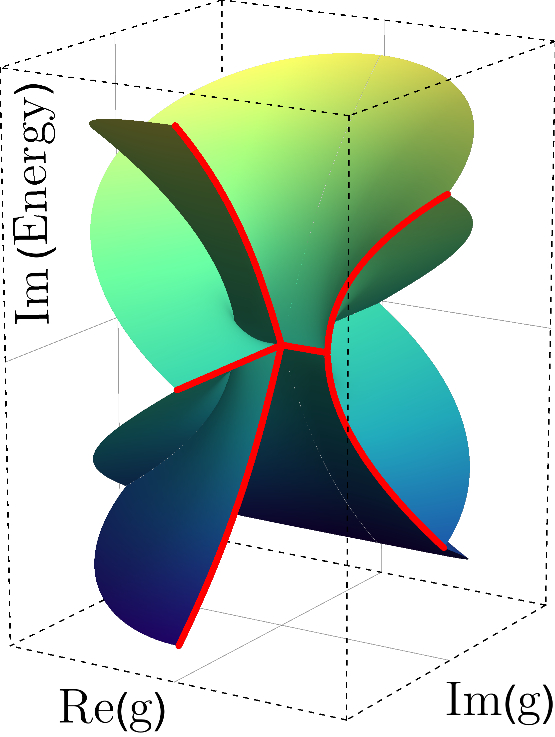}
\label{Riemann_0th_nu=omega_v4}
}
\caption[Separate sheets and complete Riemann surface of the ground 
state of the coupled oscillator system over the complex coupling plane 
($\nu=\omega$)]{Riemann surface structure of the ground-state 
energy function $E_0(g)$ over the complex coupling plane for $\nu=\omega$. 
Figures (a) and (b) show the real part of the complete surface, while (c) and 
(d) display the imaginary part. Furthermore, in Figs.~(a) and (c) the branch cuts 
are denoted as solid red lines and the decoupling limits of the energy 
$E_0(0)=\pm 2\nu \lor 0$ are denoted as red dots. 
Figures (b) and (d) show the structures with the fourth quadrant of the 
coupling plane cut out and the characteristics of the energy function  along 
the cuts denoted by solid red lines.}
\end{figure*} 
And while the vanishing coupling constant still implies that $\gamma=0$ for 
$\alpha=\beta$, as in all cases for distinct frequencies or for non-vanishing coupling, this is not the case for $\alpha=-\beta$. In that case $\gamma$ remains a free and arbitrary parameter in the wave function that has no impact on the energy eigenvalues. If $\gamma$ is not chosen to vanish the wave functions no longer 
take the separable product wave function form that we expect for the decoupled theory. As we can easily see from the relation $\alpha=-\beta=\pm \nu$ this only occurs as a special case of the decoupled theory, in which the two 
oscillators with equal frequencies are in distinct spectral phases. Nevertheless, the energy functions themselves behave as  expected when the exceptional points coalesce. We remark in particular that the coalescence of the two opposite chirality branch points leads to the formation of a simple (diabolic point) degeneracy of two states, such that the decoupled theory does not coincide with a singularity. This behaviour can for example be seen in 
Figs.~\ref{Riemann_0th_nu=omega_v1} to \ref{Riemann_0th_nu=omega_v4}, where the Riemann surface of the ground-state energy function is displayed for equal oscillator frequencies.

\section{Brief concluding remarks}

The analytic structure of two coupled oscillators that are linked via a linear coupling in both variables $x$ and $y$ has been shown to  have in general an eight-fold Riemann surface structure for the function $E_n(g)$. Regarding the eigenvalues only in the decoupling limit, one finds that these can be cast into sets of four eigenvalues, one corresponding to a \emph{conventional} phase, while the further three are associated with \emph{unconventional} phases of the system. Having identified such phases, it would be interesting to find experiments that could uncover, and perhaps even make use of these sorts of structures. First experiments that can access the complex plane or that rely on the analytic structure of the system \cite{CD, BD, DOP, XU, CHEN} give grounds for optimism.


\end{document}